\documentclass[%
reprint, 
nofootinbib,
 amsmath,amssymb,
 aps,
]{revtex4-1}
\usepackage{epsfig,epsf,rotating,wrapfig,tabularx}
\usepackage{dcolumn}
\usepackage{bm}
\usepackage{graphicx}

\begin{document}

\preprint{APS/123-QED}

\title{Hydrodynamic extension of the two component model\\ for hadroproduction  in heavy-ion collisions. }

\author{\firstname{A.~A.}~\surname{Bylinkin}}
 \email{alexander.bylinkin@desy.de}
\affiliation{%
 Institute for Theoretical and Experimental
Physics, ITEP, Moscow, Russia
}%
\author{\firstname{N.~S.}~\surname{Chernyavskaya}}
\email{nadezda.chernyavskaya@desy.de}
\affiliation{%
 Institute for Theoretical and Experimental
Physics, ITEP, Moscow, Russia
}%
\author{\firstname{A.~A.}~\surname{Rostovtsev}}
 \email{rostov@itep.ru}
\affiliation{%
 Institute for Theoretical and Experimental
Physics, ITEP, Moscow, Russia
}%

\begin{abstract}
The dependence of the spectra shape of produced charged hadrons on the size of a colliding system is discussed using a two component model. As a result, the hierarchy by the system-size in the spectra shape is observed. Next, the hydrodynamic extension of the two component model for hadroproduction using recent theoretical calculations is suggested to describe the spectra of charged particles produced in heavy-ion collisions in the full range of transverse momenta, $p_T$. Data from heavy-ion collisions measured at RHIC and LHC are analyzed using the introduced approach and are combined in terms of energy density.  The observed regularities might be explained by the formation of QGP during the collision. 
\end{abstract} 

\pacs{Valid PACS appear here}
\maketitle

\section{Introduction}
Recently, a unified approach to describe charged particle production in high-energy collisions and describing two distinct mechanisms of hadroproduction has been proposed~\cite{OUR1}. It was suggested to approximate the charged particle spectra as a function of the particle's transverse momentum $p_T$ by a sum of an exponential (Boltzmann-like) and a power law distributions:
\begin{equation}
\label{eq:exppl}
\frac{d\sigma}{p_T d p_T} = A_e\exp {(-E_{Tkin}/T_e)} +
\frac{A}{(1+\frac{p_T^2}{T^2\cdot N})^N},
\end{equation}
where  $E_{Tkin} = \sqrt{p_T^2 + M^2} - M$
with M equal to the produced hadron mass. $A_e, A, T_e, T, N$ are the free parameters to be determined by fit to the data.  

According to this approach, the exponential part stands for the release of  "thermalized" particles by the preexisting valence quarks and a quark-gluon cloud coupled to them inside the colliding baryon. The power-law term accounts for the fragmentation of  mini-jets formed by the secondary partons (gluons)
produced with a relatively large $k_T$ at the first stage of the collision,that can be described within the pQCD.  From this qualitative picture of hadroproduction one can naively expect that the spectra of charged hadrons in $\gamma \gamma$ collisions should be described by the power-law term alone due to the absence of "thermalized" quarks and gluons in the colliding systems. Such behavior has also been proven recently~\cite{OUR2}.

Thus, it is interesting to compare the shapes of charged particles produced in these two types of interactions ($\gamma \gamma$ and $pp$) with a more complex case of heavy-ion collisions.
 
\section{Hierarchy in hadroproduction dynamics}
\begin{figure*}[!ht]
\quad
\includegraphics[width =4.8cm]{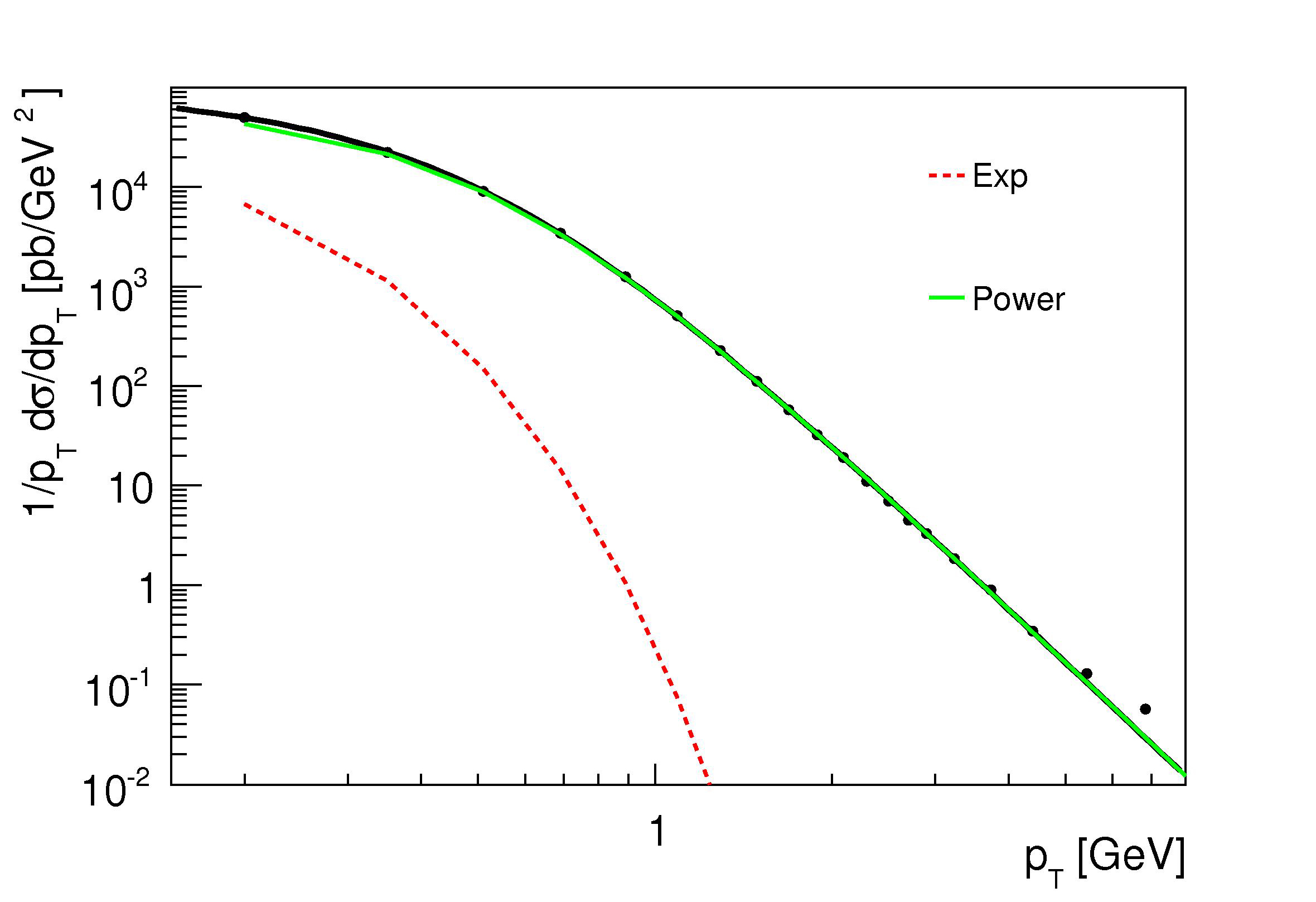}
\quad
\includegraphics[width =5cm]{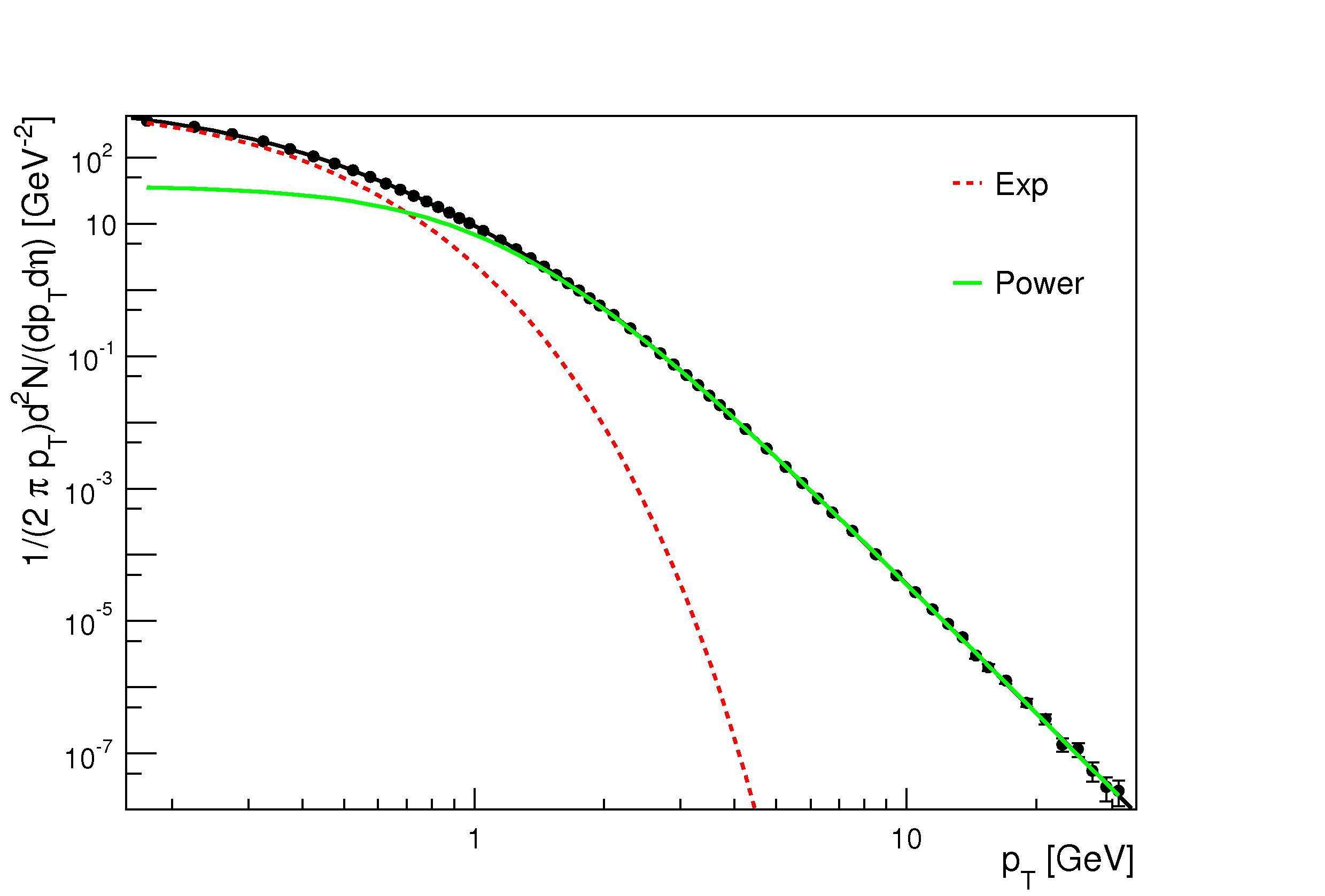}
\quad
\includegraphics[width =5cm]{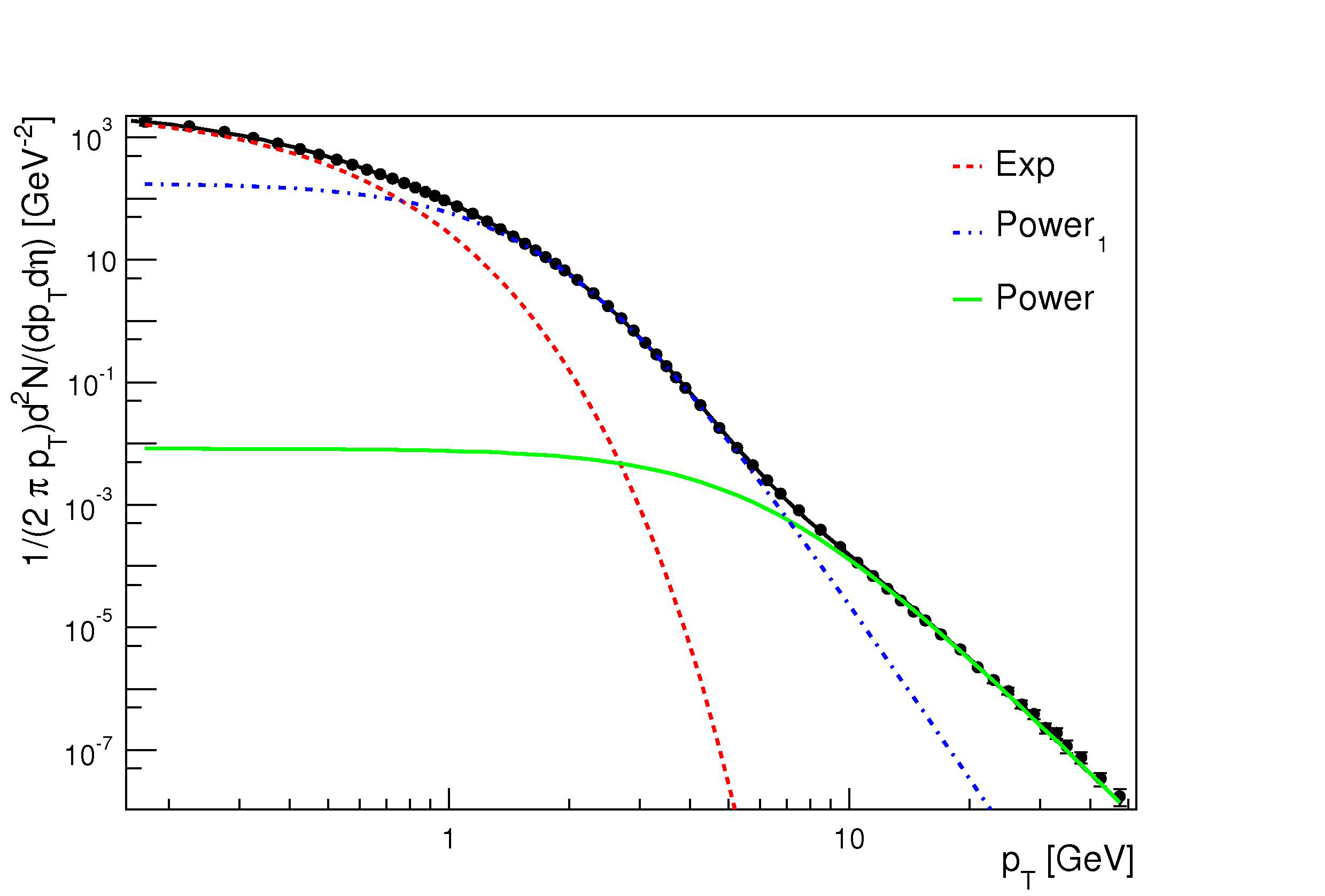}
\caption{\label{fig.1} Charged particle spectra in $\gamma \gamma$~\cite{OPAL} a), $pp$~\cite{ALICEpp} b) and in central lead-lead collisions~\cite{ALICE} c) fitted to the function~(\ref{eq:expplpl}): the red (dashed) line shows the exponential term and the green (solid) and blue (dash-dot) lines - two power-law terms.}
\end{figure*}

It is suggested to look at the recent data on lead-lead collisions measured by the ALICE Collaboration~\cite{ALICE} in the range of transverse momentum $p_T$ up to $50$ GeV.
 Figure~\ref{fig.1}  shows experimental data on $\gamma \gamma$~\cite{OPAL}, $pp$~\cite{ALICEpp} and lead-lead~\cite{ALICE} collisions fitted with the parameterization introduced~(\ref{eq:exppl}). One can notice, that this parameterization can not describe the shape of the spectra in lead-lead collisions for the very high-$p_T$ values and an additional power-law term is needed:
\begin{equation}
\label{eq:expplpl}
\frac{d\sigma}{p_T d p_T} = A_e\exp {(-E_{Tkin}/T_e)} +
\frac{A}{(1+\frac{p_T^2}{T^2\cdot N})^N} +\frac{A_{1}}{(1+\frac{p_T^2}{T^2_1\cdot N_1})^{N_1} }  
\end{equation}

\begin{figure}[!ht]
\includegraphics[width =8cm]{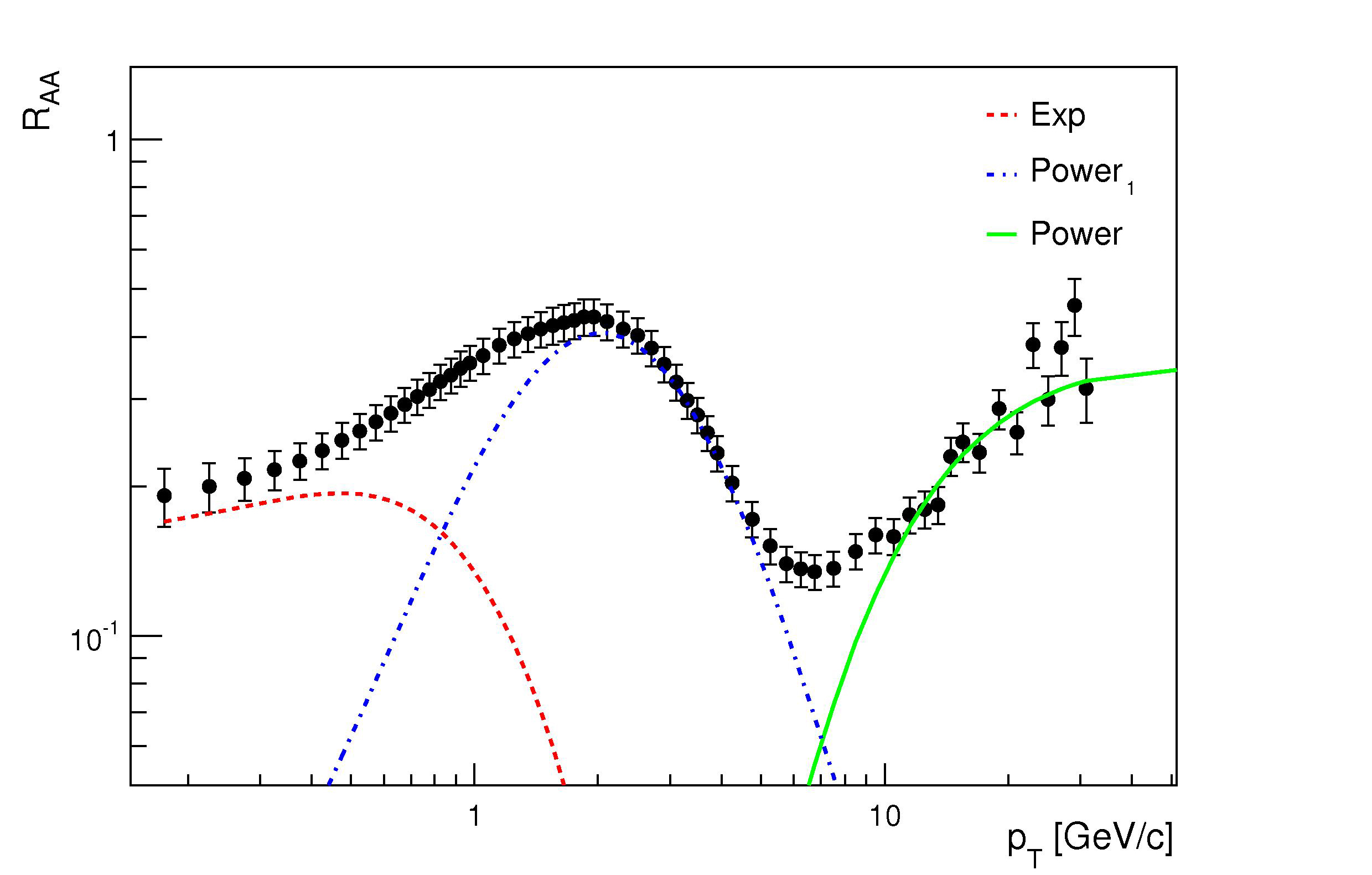}
\caption{\label{fig.2} Nuclear modification factor $R_{AA}$ measured for central lead-lead collisions~\cite{ALICE} shown together with the terms of~(\ref{eq:expplpl}) independently divided over the fit (\ref{eq:exppl}) of the $pp$-data at the same c.m.s energy: the red (dashed) line shows the exponential term and the green (solid) and blue (dash-dot) lines - two power-law terms.}
\end{figure}

Note, that an additional power-law term in lead-lead collisions might be explained by the peculiar shape of the nuclear modification factor $R_{AA}$. Figure~\ref{fig.2} shows $R_{AA}$ for lead-lead collisions measured at ALICE~\cite{ALICE} together with the lines showing contributions from the three terms of eq.~(\ref{eq:expplpl}) independently, each of them divided over the spectrum in $pp$-collisions measured at the same c.m.s. energy~\cite{ALICEpp}. One can notice, that each of these terms contribute to different regions of the transverse momentum $p_T$. The observed behavior might be explained by the following picture of hadroproduction in heavy-ion collisions:
\begin{enumerate}
\item The bulk of low-$p_T$ particles originates from the 'quark-gluon soup' formed in the heavy-ion collision and has an exponential $p_T$ distribution, as shown by the red dashed line in figures~\ref{fig.1} and \ref{fig.2}.
\item The high-$p_T$ tail (shown by the green solid line in figures~\ref{fig.1} and~\ref{fig.2}) accounts for the mini-jets that pass through the nuclei, the process that can be described in pQCD~\cite{pQCD}. When these jets hadronize into final state particles {\bf outside} the nuclei, we get the same power-law term parameter $N$ as in $pp$-collisions (figures~\ref{fig.1} b) and c)), resulting in a constant suppression ($R_{AA}$) of high-$p_T$ ($>20$ GeV) particles (figure~\ref{fig.2}). Note, that while passing through the nuclei these jets should loose about $\frac{dE}{dz}\cdot R_{A}~\sim~7~$GeV~\cite{pQCD}, where $R_{A}$ is the radius of the nuclei. Therefore, hadrons with $p_T < 7$ GeV produced from these jets will be largely suppressed, as it seen in the figure~\ref{fig.2}.
\item On the other hand, mini-jet fragmentation into final state hadrons can also occur before the jet leaves the nuclei volume. The produced particles have to wade out through the nuclei, being affected by multiple rescatterings, and thus their distribution (blue dash-dot line in figures \ref{fig.1} and \ref{fig.2})
becomes more close to the exponent, resulting in higher values of $N_1$ and $T_1$ of the power-law term, and dominates the mid-$p_T$ region. This process can't be described in pQCD, however.
\end{enumerate}

Now one can notice the hierarchy in hadroproduction dynamics by complexity (number of involved partons or initial size) of the colliding system:
\begin{itemize}
\item $\gamma \gamma $ collisions: a point like interaction that can be described in terms of pQCD and thus, need a power-law term only in its spectrum.
\item baryon-baryon collision: in addition to the mini-jet fragmentation of the virtual partons an exponential term standing for the release of thermalized particles due to preexisting quarks and gluons is added. Therefore, one gets a sum of exponential and power-law terms to describe the spectra in $pp$-collisions.
\item heavy-ion collision: due to the quenching of charged hadrons inside the nuclei the power-law term 'splits' into two distributions with different parameters (the second closer to the exponent). Therefore, we need a sum of exponential and {\bf 2} power-law terms  to describe the spectra.
\end{itemize} 
 \section{Hydrodynamic extension of the model}
Though, the parameterisation using an exponential and two power-law terms~(\ref{eq:expplpl}) gives a rather perfect description of the experimental data~\cite{ALICE} (Figure~\ref{fig.1} c), it is known that Boltzmann thermodynamics is not applicable for heavy-ion collisions. When a large colliding system is formed, one should also take effects of the 'collective motion' into account ~\cite{Hydro}. Thus, in heavy-ion collisions the multiparticle production is usually considered in terms of relativistic hydrodynamics, contrary to widely used thermodynamic approaches~\cite{Hagedorn, Tsallis} for $pp$, $\gamma p$ and $\gamma \gamma$-collisions.
 Therefore, it is suggested to modify the introduced approach ~(\ref{eq:exppl}) using recent theoretical calculations ~\cite{Hydro}.

The idea of hydrodynamic approach is that the thermalized system expands collectively in longitudinal direction generating the transverse flow by the high pressure in the colliding system. According to this approach the radiation of thermalized particles can be parameterized by the following formula: 
\begin{equation}
\label{eq:Bessel}
\frac{\mathrm{d n}}{p_{T}\mathrm{d}p_{T}}  \propto 
\int_0^R r \text{ } \mathrm{d}r \text{ }m_{T} \text{ } I_{0}\left( \displaystyle \frac{p_{T} \sinh \rho  }{T_e} \right) K_{1} \left( \frac{ m_{T}   \cosh \rho }{T_e} \right),
\end{equation}
where $\rho = tanh^{-1}\beta_r$ and $\beta_r(r) = \beta_s(\frac{r}{R})$, with $\beta_s$ standing for the surface velocity. In this analysis we take $\beta_s = 0.5 c$ which is consistent with previous observations~\cite{Hydro}.
Thus, one have to substitute the exponential term in~(\ref{eq:exppl}) by ~(\ref{eq:Bessel}).

 Note, that the power-law term in ~(\ref{eq:exppl}) stands for the point-like pQCD interactions that occur in the early stage of the collision, with the hadrons produced from the mini-jet fragmentation leave the interaction area before reaching the thermal equilibrium. Therefore, we assume this term to be considered without taking the 'collective motion' into account.

Now one can use this hydrodynamic approach to fit the recent experimental data on lead-lead collisions measured by the Alice Collaboration~\cite{ALICE} at $\sqrt{s} = 2.76~$TeV.\\

\begin{figure}[!ht]
\includegraphics[width =8cm]{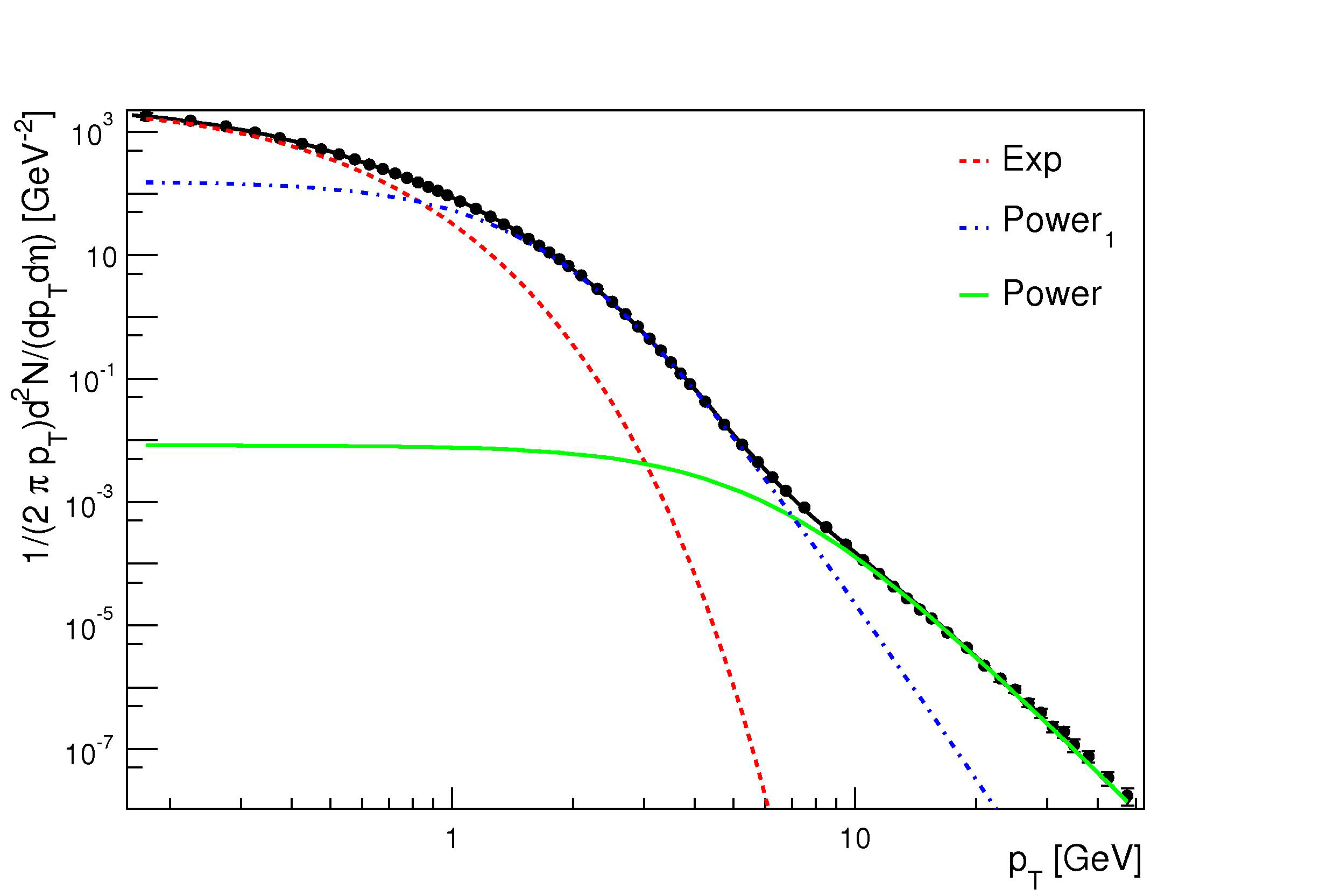}
\caption{\label{fig.3} Central lead-lead collisions~\cite{ALICE} fitted with~(\ref{eq:Besselplpl}): the red (dashed) line shows the hydrodynamic term and the green (solid) and blue (dash-dot) lines - two power-law terms.}
\end{figure}

These data are shown in figure~\ref{fig.3} together with the fit:\\

\begin{widetext}
\begin{equation}
\label{eq:Besselplpl}
\frac{\mathrm{d}n}{p_{T}\mathrm{d}p_{T}}  = A_e \cdot 
\int_0^R r \text{ } \mathrm{d}r \text{ }m_{T} \text{ } I_{0}\left( \displaystyle \frac{p_{T} \sinh \rho  }{T_e} \right) K_{1} \left( \frac{ m_{T}   \cosh \rho }{T_e} \right) + \frac{A}{(1+\frac{p_T^2}{T^2\cdot N})^N} +\frac{A_{1}}{(1+\frac{p_T^2}{T^2_1\cdot N_1})^{N_1} }.
\end{equation}
\end{widetext}

Note, that the proposed hydrodynamic extension (\ref{eq:Besselplpl}) of (\ref{eq:exppl}) only slightly modifies the description of the experimental data and still two power-law terms are needed. However, the values of the parameter $T_{e}$ extracted from parameterizations (\ref{eq:expplpl}) and (\ref{eq:Besselplpl}) differ significantly\footnote{Compare figure~\ref{fig.4} with one in \cite{OURI}.}.

\section{Freeze-out temperature\\ and combination of RHIC and LHC data}
The introduced approach (\ref{eq:Besselplpl}) allows to extract the thermalized production (described by function~(\ref{eq:Bessel})) of charged hadrons from the whole statistical ensemble.
In this paper it is proposed to study the variations of the temperature-like parameter $T_e$ in (\ref{eq:Bessel}) with the centrality and the c.m.s. energy in heavy-ion collisions. Therefore, it is interesting to consider the experimental data measured at RHIC and LHC together.

Since the centre-of-mass energies per nucleon in these experiments are varied by a factor of $\approx20$, a unified approach considering the energy density is suggested.
The energy density in heavy-ion interactions is known to depend not only on the centre-of-mass energy, but also on the centrality of the collision. Hence, while the maximum energy densities that can be reached at RHIC and at LHC differ significantly, the energy density  in central collisions at RHIC might be of the same order with that in peripheral collisions at LHC.

In this paper we consider the experimental data measured in AuAu collisions at $\sqrt{s} = 200$ GeV/N and $\sqrt{s} = 130$ GeV/N by PHENIX~\cite{PHENIX,PHENIX130} and PbPb collisions at $\sqrt{s} = 2.76$ TeV/N by ALICE~\cite{ALICE}.

The energy density $\varepsilon$ for central collisions can be determined from the experimental data by the formula~\cite{Mishustin}:
\begin{equation}
\label{eq:ed0}
\frac{dE_T}{d\eta}(\eta\sim0) = \pi R^2 \varepsilon_f \tau_0,
\end{equation}
where $\varepsilon_f$ is the energy density averaged over the transverse area, and $R$ is the nuclear radius.

However, for non-central collisions it is more convenient to estimate it using a simple parameterization~\cite{Mishustin}:
\begin{equation}
\label{eq:ed}
\varepsilon = \varepsilon_0 (\frac{s}{s_0})^{\alpha/2}{N_{coll}}^{\beta},
\end{equation}
with $\varepsilon_0$ calculated for the most central collisions, $\alpha~\approx~0.3$~\cite{Kharzeev:2001gp}, $\beta \approx 0.5$~\cite{Braun:2001us} and $\sqrt{s_0} = 200$ GeV~\cite{Mishustin}. Here the second factor is responsible for the incident energy dependence, $\sqrt{s}$ is the c.m.s collision energy, and the third one shows the dependence on the number of binary parton-parton collisions $N_{coll}$ which is related to the centrality of the collision. Note, that in this analysis $\varepsilon_0$ turned out to be the same for PHENIX and ALICE data, thus confirming the $\alpha = 0.3$ value proposed in \cite{Kharzeev:2001gp}.

\begin{figure}[!ht]
\includegraphics[width =8cm]{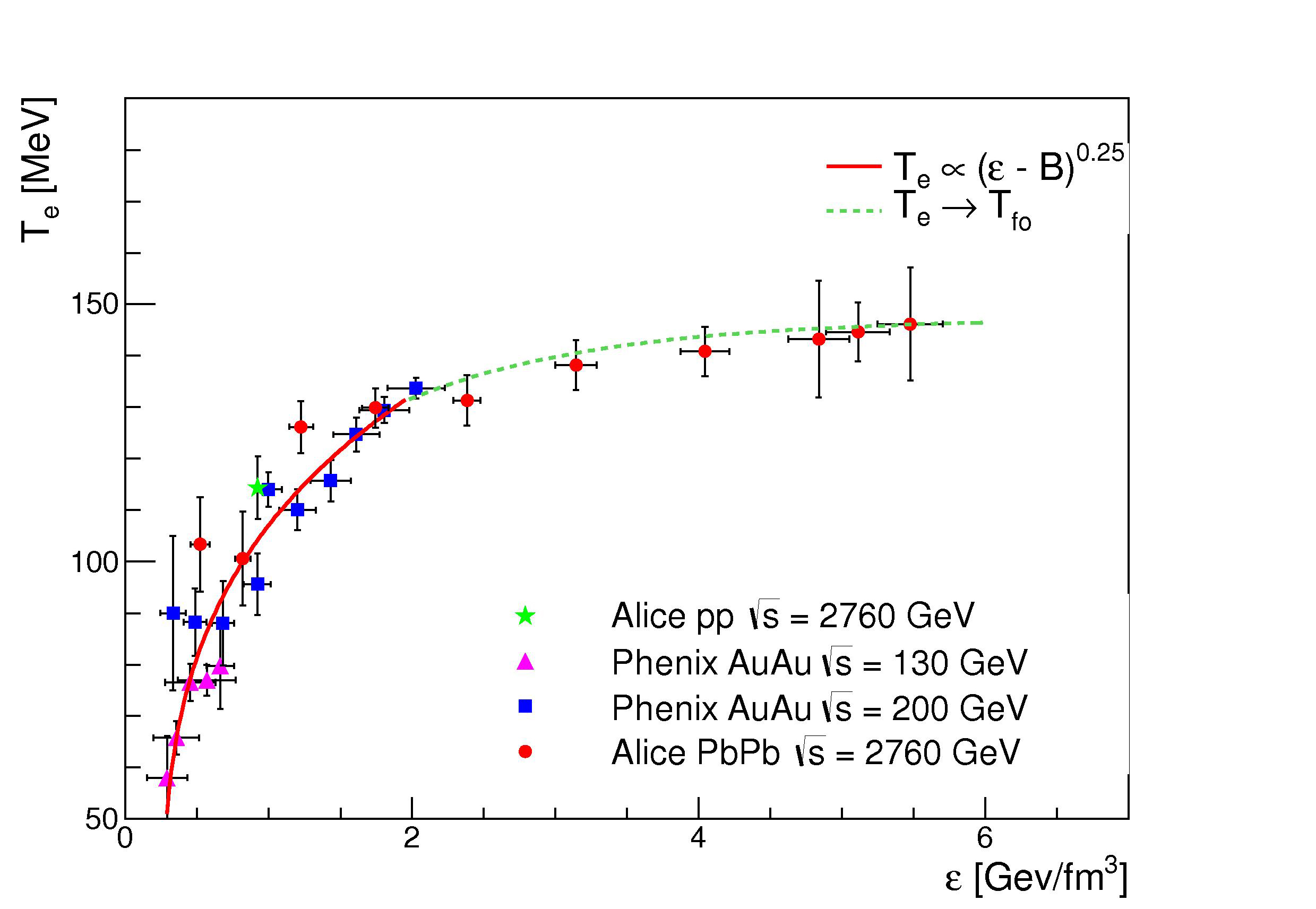}
\caption{\label{fig.4} Temperature of the final state hadrons coming from the 'thermalized' part of the spectra in heavy-ion collisions as a function of energy density. Solid line stands for the $T_e~\propto~(\varepsilon-B)^{0.25}$ fit and dashed line shows $T_e\rightarrow const$ behavior.}
\end{figure}

Having calculated the energy density $\varepsilon$ using the formula~(\ref{eq:ed}), one can plot the temperature $T_e$ extracted from (\ref{eq:Besselplpl}) as a function of it, as shown in figure ~\ref{fig.4}.
First of all, as it was expected, the energy density obtained in central collisions at RHIC is similar to those in peripheral collisions at LHC, and, remarkably, a smooth transition in the $T_e$ values between these three measurements is also observed. Note, that as one could naively expect, the value of $T_e$ (as well as $N$ and $T$ of the power-law term in (\ref{eq:Besselplpl})) for peripheral lead-lead collisions turns out to be practically identical with that obtained for $pp$-collisions at the same c.m.s. energy~\cite{ALICEpp}.

Next, one can notice rather interesting behavior of the temperature $T_e$ as a function of energy density ($\varepsilon~\propto~T_{e}^{4}~+~B$), which is in a good agreement with the Bag model~\cite{Bag}, with $B = 0.25$ GeV/fm$^{3}$, as determined from the fit in figure~\ref{fig.4}.
 Another remarkable observation on the temperature $T_e$ of the final state particles is that for high energy densities it reaches a certain limit. This might be explained from QGP theory that considers the phase transition temperature $T_{c}$ from QGP to final state hadrons: the expanding system cools down until it reaches the freeze-out stage, thus, the temperature of the final state particles should be always below $T_{c}$.  Indeed, for high values of $\varepsilon$ one can notice, that the observed freeze-out temperature is $T_{fo}\approx 145$ MeV, and (as one can expect) is slightly below the critical temperature $T_{c}\sim155-160$ MeV for QGP obtained in different calculations~\cite{QGP1,QGP2}.\\

\section{Conclusion}
The spectra of charged hadron production in heavy-ion collisions have been compared with those measured in $pp$ and $\gamma \gamma$ interactions using the recently introduced two component model. The observed hierarchy on the size of the colliding system has been discussed and the qualitative picture for hadroproduction in heavy-ion collisions explaining the peculiar shape of nuclear modification factor, $R_{AA}$, has been introduced. Next, the hydrodynamic extension of this parameterization accounting for the collective motion in heavy-ion collisions was suggested. This approach allowed to extract the 'thermalized' production of charged hadrons from the whole statistical ensemble and to study it separately. Thus, the variations of the temperature of the final state hadrons coming from the 'thermalized' part of the spectra have been studied as a function of energy density using both RHIC and LHC data and the behavior that might be explained in terms of QGP formation has been observed.\\

\begin{acknowledgements}
The authors thank Professor Mikhail Ryskin for fruitful discussions and his help provided during the preparation of this paper. 
\end{acknowledgements}


\end{document}